\documentstyle[aps,pra,epsf]{revtex}
\input{epsf}
\begin{document} 
\title{The asymmetric lossy near-perfect lens} 
\author{S. Anantha Ramakrishna\footnote{E-mail: s.a.ramakrishna@ic.ac.uk,
Tel: +44 020 75947597, FAX: +44 020 75947604}, J.B. Pendry\footnote{E-mail:
j.pendry@ic.ac.uk}}
\address{The Blackett laboratory, Imperial College, London SW7 2BZ, UK}
\author{D. Schurig, D.R. Smith\footnote{E-mail: drs@ucsd.edu} and S. Schultz}
\address{Department of Physics, University of California, San Diego, 9500
Gilman Drive, La Jolla, CA 92093-0319, USA}
\maketitle 
\begin{abstract}
We extend the ideas of the recently proposed perfect lens [J.B.  Pendry, Phys.
Rev.  Lett.  {\bf 85}, 3966 (2000)] to an alternative structure.  We show that
a slab of a medium with negative refractive index bounded by media of different
positive refractive index also amplifies evanescent waves and can act as a 
near-perfect lens. We examine the role of the surface states 
in the amplification of the evanescent waves. 
The image resolution obtained by this asymmetric lens is more
robust against the effects of absorption in the lens.  In particular, we study
the case of a slab of silver, which has negative dielectric constant, with air
on one side and other media such as glass or GaAs on the other side as an
`asymmetric' lossy near-perfect lens for P-polarized waves.  It is found that
retardation has an adverse effect on the imaging due to the positive
magnetic permeability of silver, but we conclude that subwavelength image
resolution is possible inspite of it.

\end{abstract}

%\newpage
\section{Introduction} The electromagnetic radiation emitted or scattered by an
object consists of a radiative component of propagating modes and a near-field
component of non-propagating modes whose amplitudes decay exponentially with
distance from the source.  For a monochromatic source, the electromagnetic
field in free space can be expressed as a Fourier sum over all wave vectors:
\begin{equation} 
E(x,y,z;t) = \sum_{k_{x},k_{y},k_{z}} E(k_{x},k_{y},k_{z}) \exp[i(k_{x}x+k_{y} 
y+
k_{z}z-\omega t)] 
\end{equation} 
where $k_{x}^{2}+k_{y}^{2}+k_{z}^{2} =\omega^{2}/c^{2}$ and $c$ is the speed of
light in free space.  For $k_{x}^{2} +k_{y}^{2} > \omega^{2}/c^{2}$, $k_{z}$ is
seen to be purely imaginary, and the case is similiar for $k_{x}$ and $k_{y}$.
The near-field consists of these partial waves with imaginary wave vectors and
decays exponentially away from the source.  These are the high-frequency
Fourier components describing the fast varying spatial 
features on the object and are
never detected in conventional imaging using conventional lenses.  This lack of
information results in the limitation to conventional imaging that
sub-wavelength features of a source cannot be resolved in the image.  In order
to overcome this limitation, scanning near-field optical microscopy (SNOM) was
proposed where the near-field of the radiating object is probed by bringing a
tapered fibre tip very close to the object (See Ref.~\cite{greffet&carminati}
for a recent review).  A rule of the thumb is that the near-field produced by a
(periodic) feature of spatial extent $d$ will decay exponentially at a rate of $d/2\pi$ away
from the surface.  Hence there is a need to get close to the surface at these
resolutions, just in order to detect the evanescent field.  
Impressive advances have been made in this field, but there still
remain several problems in understanding the images, phase contrast mechanisms
and associated artifacts.

Now in order to completely reconstruct the object, we would need to amplify
these evanescent waves and give the appropriate phase shift for the propagating
components.  This is precisely what the recently proposed perfect lens with
negative refractive index accomplishes\cite{pendry00}.  Veselago had noted a
long time ago\cite{veselago} that due to the reversal of Snell's law, a slab of
negative refractive index would act as a lens in that the rays from a source on
one side would get refocussed on the other side to form an image.  But the
amplification of evanescent waves by a slab of negative refractive index noted
by Pendry\cite{pendry00} was a surprising and new result.  The possibility of negative
refractive media has already been experimentally demonstrated in the microwave
region of the spectrum\cite{smith00,smith01}.  Both the dielectric constant and 
the
magnetic permeability are negative in a negative refractive medium, and the
electromagnetic waves in such a medium will be left-handed as a consequence of
Maxwell's equations.  No known natural materials have negative magnetic
permeability, and the negative refractive media are meta-materials consisting
of interlaced periodic arrays of split ring resonators\cite{pendryIEEE} and
thin wires\cite{pendry96}.  Pendry also showed that a thin slab of silver with
negative dielectric constant would act as near-perfect lens for near-field
imaging with P-polarized waves in the electro-static limit, with the resolution
limited only by absorption in the lens (silver).  More recently, we have
examined some consequences of deviations in the dielectric constant and
magnetic permeability from the perfect lens conditions ($\epsilon = -1$, $\mu =
-1$) on the resolution of the lens \cite{smithprep} and found that
the restrictions on $\epsilon, \mu$ were quite severe, but achieveable by
current day technology. The main advantage of this near-field perfect lens over 
the current methods of SNOM is that a complete image is generated at the image 
plane. This would be important in many applications, for example, in-situ 
imaging of biological molecules and processes by flourescence imaging.  

Here we extend the ideas of Pendry's original work.  We show that an asymmetric
lens consisting of a slab of negative refractive index bounded by media of
different positive refractive index can also act as a near-perfect lens.  We then
examine the nature of the surface modes which are responsible for the
amplification of the evanescent waves.  In particular, we study the case of a
film of silver (with negative dielectric constant) deposited on other media
such as glass or GaAs with positive dielectric constant as a near-field imaging
lens at optical frequencies.  We show that the asymmetry in the system can
actually enhance the resolution, depending on the choice of asymmetry and the
operating frequency.  One of the main advantages of this asymmetric lens is
that a very thin metal film deposited on a solid substrate (Glass or GaAs) will 
be mechanically much more stable than a free-standing metal film.  We also 
study the effects of retardation on the system and conclude that
sub-wavelength resolution is possible inspite of the adverse effects of
retardation and absorption.

\section{The asymmetric slab}  
[Insert figure 1 about here ]\\

Typically we consider the asymmetric lens to be a slab of negative refractive
medium of thickness $d$, dielectric constant $\epsilon_{2}$ and magnetic
permeability ($\mu_{2}$) between media of differing refractive indices (See
figure~1).  We will consider the medium-1 on one side of the source to be air
($\epsilon_{1} =1$, $\mu_{1} = 1$) and some other dielectric or magnetic
material ($\epsilon_{3}$, $\mu_{3}$) on the other side.  We will consider the
object plane to be in medium-1 at a distance $d/2$ and the image plane, where
we detect the image inside
 medium-3 at a distance $d/2$ from the edge of the slab.
  Let  P-polarized light be incident on the
slab from the medium-1, with the magnetic field\footnote{It is more convenient
to use the magnetic field for the P-polarized light .  The electric field can
be obtained by using the Maxwell's equation  $\vec{k}^{(j)} \times \vec{B} = 
-\omega 
\epsilon_{j} \vec{E}$.} given by
\begin{equation} 
H_{1p} = \exp(ik_{z}^{(1)}z + ik_{x} x -i\omega t), 
\end{equation} 
where $k_{x}^{2} + {k_{z}^{(j)}}^{2} = \epsilon_{j} \mu \omega^{2}/c^{2}$ ($j = 
1,2,3$ for the different media). We will work in two dimensions for reasons of 
simplicity.  If the index of refraction is negative ($\epsilon_{2} <0$, 
$\mu_{2} <0$), then the Maxwells equations and causality demand that $k_z^{(2)} 
= - \sqrt{ \epsilon_{2} \mu_{2} \omega^{2}/c^{2} - k_{x}^{2}}$ if $k_{x} < 
\sqrt{\epsilon_{2}\mu_{2}} \omega /c$, and $k_z^{(2)} = i \sqrt{ k_{x}^{2} - 
\epsilon_{2} \mu_{2} \omega^{2}/c^{2}}$ if $k_{x} > \sqrt{\epsilon_{2}\mu_{2}} 
\omega /c$ \cite{pendry00}.  
The transmission coefficient for the P-polarized light across the slab is 
\begin{equation} 
T_{p}(k_{x}) = \frac{ 4 
~(\frac{k_{z}^{(1)}}{\epsilon_{1}})~(\frac{k_{z}^{(2)}}{\epsilon_{2}}) 
~\exp(ik_{z}^{(2)} d) }{ ( \frac{k_{z}^{(1)}}{\epsilon_{1}} + 
\frac{k_{z}^{(2)}}{\epsilon_{2}})(\frac{k_{z}^{(2)}}{\epsilon_{2}} + 
\frac{k_{z}^{(3)}}{\epsilon_{3}}) - (\frac{k_{z}^{(1)}}{\epsilon_{1}} - 
\frac{k_{z}^{(2)}}{\epsilon_{2}})(\frac{k_{z}^{(3)}}{\epsilon_{3}} - 
\frac{k_{z}^{(2)}}{\epsilon_{2}})\exp (2ik_{z}^{(2)}d) } .  
\end{equation} 
We immediately see that if either $\epsilon_{2} = -\epsilon_{1}$ and $\mu_{2} =
-\mu_{1}$, or $\epsilon_{2} = -\epsilon_{3}$ and $\mu_{2} = -\mu_{3}$, then
$T_{p} = [2 \epsilon_{3}k_{z}^{(1)}/(\epsilon_{1}k_{z}^{(3)}+\epsilon_{3}k_{z}^{(1)})] \exp (-ik_{z}^{(2)} d)$
and the amplification of the evanescent waves as well as the phase reversal for
the propagating components results.  Thus, the system does not have to be
symmetric ($\epsilon_{1} = \epsilon_{3}$, $\mu_{1}=\mu_{3}$) as in the original
work of Pendry for accomplishing amplification of evanescent waves. In this
asymmetric case, however, the field 
strength at the image plane differs from the object plane by a
constant factor, and thus the image intensity is changed. Further, it is 
easily verified that wave-vectors with different $k_x$ will refocus at 
slightly different positions unless $\epsilon_{3}=\epsilon_{1}$ and $\mu_{3}
=\mu_{1}$ (i.e. the symmetric case) or in the limit $k_{x} \rightarrow \infty$.
Thus there is no unique (perfect) image plane and we should expect the image to
suffer from aberrations. Hence, we term the asymmetric slab {\it a near-perfect
lens}. 
A similiar result holds for the S-polarized wave incident on the slab.
Again, we point out that with this asymmetric lens, the object is considered 
to be in medium-1 (usually air) and the image is formed inside medium-3 (mostly
considered to be a high index dielectric in subsequent sections of this paper).

In the electrostatic (magnetostatic) approximation ($k_{x} \rightarrow
\infty$), we have $k_{z}^{(j)} \rightarrow ik_{x}$ and the dependence of the
transmission coefficient for the P(S)-polarized field on $\mu$($\epsilon$) is
eliminated.  The transmission coefficient for the P-polarized wave across the
slab is then given by 
\begin{equation} 
T_{p}(k_{x}) = \frac{ 4
\epsilon_{2}\epsilon_{3}\exp (-k_{x}d) }{ (\epsilon_{1}+
\epsilon_{2})(\epsilon_{2}+\epsilon_{3}) - (\epsilon_{2}-\epsilon_{1})
(\epsilon_{2}-\epsilon_{3})\exp (-2k_{x}d) } , 
\end{equation} 
and the amplification of the evanescent waves depends only on the condition on
the dielectric constant ($\epsilon_{2} = - \epsilon_{1}$ or $\epsilon_{2} = -
\epsilon_{3}$).  Such a system can be easily realized as several metals act as a
good plasma in some range of optical frequencies in that they can have a large
negative real part of the dielectric constant with a comparatively small
imaginary part.  Typically, we take our system to be a thin film of silver
deposited on another medium such as glass, GaAs, Silicon.  We note that silver
is highly dispersive and by choosing the operating wavelength of light
approprately, the dielectric constant ($\epsilon_{2}$) of silver can be chosen
to be either $-\epsilon_{1}$ or $-\epsilon_{3}$.  Note that $\mu = +1$
everywhere for such a system.  In the electrostatic limit, we can take the
object plane and the image plane to be symmetric about the slab at distance a
$d/2$ from the edge of the slab. To arrive at a simple, though approximate, 
description of the asymmetric lens, we will take the electrostatic limit
in the remainder of this section. We will consider the effects of retardation
in Section IV.  

	The link between the amplification of the evanescent waves to the
presence of a surface plasmon mode has already been pointed out\cite{pendry00}.
To obtain an insight into the  process, let us examine the spatial field 
variation in our system. First, let us consider the transmission ($T$) and the 
reflection ($R$) from the slab as a sum of partial waves arising from multiple 
scattering at the interfaces:
\begin{eqnarray}
T &=& t_{21} t_{32} e^{-k_{x}d} + t_{21} r_{32} r_{12}t_{32} e^{-3k_{x}d} +
t_{21} r_{32} r_{12}r_{32} r_{12}t_{32} e^{-5k_{x}d} + \cdots~~, \nonumber \\
 &=& \frac{ t_{21} t_{32} e^{-k_{x}d} }{1-r_{32} r_{12}e^{-2k_{x}d}}, \\
R &=& r_{21} + t_{21} r_{32} t_{12} e^{-2k_{x}d} + t_{21} r_{32} r_{12}r_{32} 
t_{12} e^{-4k_{x}d} + \cdots ~~,\nonumber \\
 &=& r_{21} + \frac{t_{21} r_{32} t_{12} e^{-2k_{x}d}}{1-r_{32} 
r_{12}e^{-2k_{x}d}} ,
\end{eqnarray}
where $t_{jk} = 2\epsilon_{j}/(\epsilon_{k}+\epsilon_{j}) $ and $r_{jk} = 
(\epsilon_{j} -\epsilon_{k})/(\epsilon_{k}+\epsilon_{j}) $ are the partial 
transmission and reflection Fresnel coefficients for P-polarized light
(obtained by matching the 
tangential components of $E$ and $H$) across the interface between Media-(j) 
and (k) (see figure~1). When the perfect-lens condition at any of the interfaces 
is satisfied, the partial reflection coefficient $r_{jk}$ as well as the 
partial transmission coefficient $t_{jk}$ for  evanescence across the interface 
diverges.  It should be noted, however, that the S-matrix that relates the 
incoming and the outgoing wave amplitudes in the scattering process is analytic 
in the complex (momentum or energy) plane. Hence, 
the  sum of the infinite series is still valid due to 
the analyticity of the S-matrix, and 
the reflection and the transmission from the slab are well-defined quantities.  
The above procedure is entirely equivalent to the usual time-honoured practice of 
decomposing the fields into a complete set of basis states and matching the 
field amplitudes at the boundaries as follows.
Let the fields in regions -- 1, 2, and 3 be given by 
\begin{eqnarray}
H_{1} &=& e^{-k_{x}z} + R e^{k_{x}z}, \\ 
H_{2} &=& A e^{-k_{x}z} + Be^{k_{x}z},\\ 
H_{3} &=& T e^{-k_{x}z}, 
\end{eqnarray} 
where $R$ and $T$ represent the reflection and transmission coefficients of the 
slab. Matching the tangential components of $H$ and $E$ at the boundaries, we 
obtain after some trivial  algebra:
\begin{eqnarray} 
A(\epsilon_{2} = -\epsilon_{1}) =
\left(\frac{\epsilon_{3}-\epsilon_{1}} {\epsilon_{3}+\epsilon_{1}}\right)
e^{2k_{x}d},~~~~ A(\epsilon_{2} = -\epsilon_{3}) = 0 ,\\ 
B(\epsilon_{2} = -\epsilon_{1}) = 1,~~~~ B(\epsilon_{2} = -\epsilon_{3}) = 
\frac{2\epsilon_{3}}{\epsilon_{3}+\epsilon_{1}} ,\\ 
T(\epsilon_{2} = -\epsilon_{1}) = \frac{2 \epsilon_{3}}{\epsilon_{3}+
\epsilon_{1}} e^{k_{x}d},~~~~T(\epsilon_{2} = -\epsilon_{3}) =\frac{2
\epsilon_{3}}{\epsilon_{3}+\epsilon_{1}}e^{k_{x}d} ,\\ 
R(\epsilon_{2} =-\epsilon_{1}) = \left(\frac{\epsilon_{3}-\epsilon_{1}}
{\epsilon_{3}+\epsilon_{1}}\right) e^{+2k_{x}d},~~~~ R(\epsilon_{2} =
-\epsilon_{3}) = \frac{\epsilon_{3}-\epsilon_{1}}{\epsilon_{3}+\epsilon_{1}}.
\end{eqnarray} 
Note that there are two separate cases, $\epsilon_{2} = -\epsilon_{1}$ on the 
left side, and $\epsilon_{2} = -\epsilon_{3}$ on the right side in the above
equations. 
 
[Insert figure 2 about here ]\\

First, we note that  while the transmission is the same in  both the 
cases, the spatial variation of the field is completely different as 
shown in figure~2. 
Second, there is a non-zero reflection in both cases, and for $\epsilon_{2} = 
-\epsilon_{1}$ the reflection is also amplified. 
The first case of $\epsilon_{2} = -\epsilon_{1}$ is exactly the condition for
the excitation of a surface plasmon state at the air-silver interface, where the
field is seen to be large and decaying on either side of the interface.  
The other case of
$\epsilon_{2} = -\epsilon_{3}$ corresponds to the excitation of a surface
plasmon state at the silver-GaAs interface.  In this case only the growing
solution within the silver slab is present.  It is interesting to note that in
the symmetric case of $\epsilon_{1} =\epsilon_{3}$, again only this growing
solution within the slab is selectively excited. The reflectivity 
is also zero for the symmetric slab. To understand the role of the surface 
modes in the amplification of the evanescent field, we note that the condition
$\epsilon_{2} = -\epsilon_{1,3}$ is exactly that for a surface plasmon state
to exist  on an interface between the semi-infinite negative and semi-infinite 
positive media. In a finite slab with two surfaces, 
however, the surface plasmons  detune from the resonant frequency, and the
detuning is exactly of the right magnitude to ensure that they are excited 
by incident fields to the correct degree for a focussed image.  
 
The presence of a large reflectivity has 
serious consequences for the use of a lens for near-field imaging applications 
as it would disturb the object field. We call the first case of $\epsilon_{2} = 
-\epsilon_{1}$, where the reflection is also amplified, as the unfavourable 
configuration, and the second case of $\epsilon_{2} = -\epsilon_{3}$ as the 
favourable configuration of the asymmetric lens. Obviously it would be most 
advantageous if the reflectivity were zero. Here we draw an analogy with a 
conventional lens which in the ideal has perfectly transmitting surfaces, 
but in practice always produces some stray reflection from the lens surface, 
but nevertheless still provides an acceptable degree of functionality. 

\section{Effects of absorption}

 Now we will consider the effect of absorption in the slab of the negative
index medium on the imaging process.  We will work in the electrostatic limit
or assume that the perfect lens conditions for the magnetic permeability
($\mu_{2}=-\mu_{1}$ or $\mu_{2}=-\mu_{3}$) is also satisfied.  Pendry
\cite{pendry00} had shown that the eventual resolution is limited by absorption
in the lens.  This absorption can be included in the calculation by adding an
imaginary part to the dielectric constant.  Pendry had used the dispersion
$\epsilon_{2}(\omega) = 5.7 - 9.0^{2}(\hbar\omega)^{-2} + i 0.4$ for silver
($\hbar \omega$ in eV) and we will continue to use it here.  Put $\epsilon_{2}
= -\epsilon_{k} + i \epsilon_{2}'$, where $k$ is either $1$ or $3$.  The
expression for the transmission becomes (assuming $\epsilon_{2}' \ll
\epsilon_{k}$),
\begin{equation} 
T_{p} = \frac{- 4 \epsilon_{k}\epsilon_{3}\exp
(-k_{x}d) }{ \pm i\epsilon_{2}'(\epsilon_{1}-\epsilon_{3} ) - 2\epsilon_{k}
(\epsilon_{1}+\epsilon_{3}) \exp (-2k_{x}d) } 
\end{equation}
 with the $\pm$ being chosen according to $k$ is either $1$ or $3$
respectively.  The resolution in Pendry's original case was limited because for 
large $k_{x}$ the exponential term in the denominator becomes smaller than the 
other term.  Now, we can see that there is definitely an advantage in
choosing $\epsilon_{k}$ to be the larger of $\epsilon_{1}$ and $\epsilon_{3}$,
in order to make the term containing the exponential to dominate in the
denominator.  Thus the asymmetry can actually help us better the limit on the
resolution set by absorption.  Fortunately, this also corresponds to the
favourable case of $\epsilon_{2} = -\epsilon_{3}$.  Choosing $\epsilon_{k}$ to
be the smaller of the two would, in contrast, cause degradation of the
resolution.  Hence, the sub-wavelength resolution proposed earlier by us as the
ratio of the optical wavelength to the linear size of smallest resolved feature
\cite{smithprep} works out to be
\begin{equation} 
res =\lambda_{0}/\lambda_{min} = -\frac{\ln \vert \epsilon_{2}'/2 
\epsilon_{3}\vert\lambda_{0}}{4 \pi d}, 
\end{equation}1
(assuming $\epsilon_{3} \gg \epsilon_{2}'$ and$\epsilon_{3} \gg \epsilon_{1}$) 
in the favourable configuration.

[Insert figure 3 about here ]\\

Below we will consider the above effects of absorption on the image of two
uniformly illuminated slits (intensity =1) obtained in transmission by the
asymmetric slab of silver of thickness $d=$40nm.  On the other side, we
consider several media of higher dielectric constant.  We numerically obtain
the image at the image plane (considered to be at $d/2$ in the electro-static
limit) and plot these in figures~3(a, b).  
In figure~3(a), we consider the
case where the resolution is enhanced due to the asymmetry in the lens.  We
take $ \epsilon_{2} = -\epsilon_{3} + i \epsilon_{2}'$ by tuning to the
appropriate frequency (lower, according to the dispersion form), where
$\epsilon_{3} > 1$ for all the media considered:  water ($\epsilon_{3} = 1.77$),
glass ($\epsilon_{3} = 2.25$), zirconia ($\epsilon_{3} = 4.6$) and silicon
($\epsilon_{3} = 14.9$ at $\hbar \omega_{sp} = 2 eV$ corresponding to the
surface plasmon being excited at the silver-silicon interface).  The level of
absorption in silver is taken to be approximately the same at all these
frequencies.  We consider the object to consist of two slits, 20nm wide and
placed apart by a centre to centre distance of 80nm.  We find that the best
contrast is obtained for the case of silicon rather than the symmetric lens.
In fact, the resolution initially degrades with increasing $\epsilon_{3}$, but
improves drastically for large values of $\epsilon_{3}$.  In this case, the two
slits are not resolved for the case of glass and zirconia.  We obtain a limit of the
spatial resolution of about 70nm (centre to centre) for the case of silicon.
In figure~3(b), we consider the images
obtained for the case $\epsilon_{2} = -\epsilon_{1} + i \epsilon_{2}'$, where
$\epsilon_{1} = 1.0$ for air on one side.  This is the unfavourable
configuration corresponding to  the surface plasmon excitation 
($\hbar \omega_{sp} = 3.48 eV$)
at the air-silver interface.  We consider a larger object of two slits
of width 30nm, the centres placed apart by 120nm and find that the image
resolution gets degraded with increasing $\epsilon_{3}$.  For the case of
Silicon ($\epsilon_{3} = 16.4$ at this frequency), the image is just resolved
according to the Rayleigh criterion\footnote{It should be noted that the 
Rayleigh criterion, in the strict sense,
 is applicable for the radiative modes. It states that 
two separate sources are just resolved when the principal maximum of one image 
coincides with the first minimum in the diffraction pattern of the other. This 
results in the ratio of the intensity at the mid-point between the sources to 
the intensity at the maxima of $\simeq 0.811$. We will use this ratio as the 
equivalent measure for defining the resolution in our case} for resolution.  
However, we will not be particularly interested in this case due to the 
amplified reflected wave which will severely affect the imaging.

\section{Effects of retardation}

In the electrostatic(magnetostatic) limit of large $k_{x} \sim k_{z} \sim
q_{z}$, there is no effect of changing $\mu$($\epsilon$) for the
P(S)-polarization.  The deviation from the electrostatic limit caused by the
non-zero frequency of the electromagnetic wave would, however, not allow this
decoupling.  We will now proceed to investigate the effects of retardation
caused by a finite frequency of the light.  It has already been
noted\cite{smithprep} that a mismatch in the $\epsilon$ and $\mu$ from the
perfect-lens conditions would always limit the image resolution and also leads
to large transmission resonances associated with the excitation of coupled
surface modes that could introduce artifacts into the image.  

[Insert figure 4 about here ] \\

Let us first examine the nature of the surface plasmons (for P-polarized 
incident light) at a single interface between a positive and a negative medium. 
The condition for the existence of a surface plasmon at the interface is 
\cite{raether}
\begin{equation}
\frac{k_{z}^{(1)}}{\epsilon_{1}}  + \frac{k_{z}^{(2)}}{\epsilon_{2}} = 0,
\label{singleplasmon}
\end{equation}
which gives the dispersion relation:
\begin{equation}
k_{x}= \frac{\omega}{c} \left[ \frac{\epsilon_{2}(\epsilon_{2}-\mu_{2})}{ 
\epsilon_{2}^{2}-1} \right]^{\frac{1}{2}},
\end{equation}
where it is assumed $\epsilon_{1} =1$ and $\mu_{1}=1$ for the positive medium. 
Note that Eq.~(\ref{singleplasmon}) can be satisfied only for imaginary 
$k_{z}^{(2)}$, when $\epsilon_2$ is negative. 
The dispersion  is plotted in figure~4, where the causal 
plasma dispersion forms $\epsilon_{2} =1 - \omega_{p}^{2}/\omega^{2}$ and 
$\mu_{2}\simeq 1 - \omega_{mp}^{2}/\omega^{2}$ are assumed for the material of 
the negative medium\cite{pendry96,pendryIEEE}. This is because a  
dispersionless $\mu(\omega) <0$ 
at all $\omega$ would not be physical\cite{landau}. We can see that the plasmon 
dispersion take different forms for $\omega_{mp} > \omega_{p}$ (the upper curve) 
and $\omega_{mp} < \omega_{p}$ (the lower curve). At large $k_x$, the plasmon 
frequency tends to the electrostatic limit of $\omega_{p}/\sqrt{2}$ from either 
above ($\omega_{mp} > \omega_{p}$ ) or (below $\omega_{mp} < \omega_{p}$). At 
small $k_x$, the surface plasmon first appears at the light-line ($\omega 
=ck_{x}$). We note that similiar results have been obtained by Ruppin very 
recently for the dispersion of the surface plasmon modes in negative refractive 
media for the case $\omega_{p}>\omega_{mp}$\cite{ruppin}.

[Insert figure 5 about here ] \\

Now, let us next consider the  symmetric lossless slab ($\epsilon_{1} =
\epsilon_{3}$, $\mu_{1}=\mu_{3}=1$, $\epsilon_{2}' = 0$) and examine the nature
of these surface modes.  As is well known, the two degenerate surface plasmons 
at the two interfaces get coupled for a thin slab and gives rise to coupled slab 
modes: a symmetric and an antisymmetric mode. 
The condition for these resonances (for P-polarized light) are \cite{raether}
\begin{eqnarray}
\tanh(k_{z}^{(2)} d /2) = -\epsilon_{2} k_{z}^{(1)}/ k_{z}^{(2)} ,\\
\coth(k_{z}^{(2)} d /2) = -\epsilon_{2} k_{z}^{(1)}/ k_{z}^{(2)} ,
\label{resonancecond} 
\end{eqnarray}
where $k_{z}^{(j)} = \sqrt{k_{x}^{2}-\epsilon_{j}\mu_{j}\omega^{2}/c^{2}}$.
Using the above conditions, the dispersion relation for the coupled surface
modes can be obtained and we show them in figure~4, where again we have assumed
the dispersion forms $\epsilon (\omega) = 1-\omega_{p}^{2}/\omega^{2} $ and 
$\mu_{2}= 1 - \omega_{mp}^{2}/\omega^{2}$.  We find
that the finite frequency changes the dispersion from the electrostatic limit
(where it is given by $\omega = \omega_{p}\sqrt{1 \pm \exp(-k_{x}
d)}/\sqrt{2}$).  The dispersion relations are qualitatively different for
$\omega_{mp} < \omega_{p}$ (or $\mu_{2}(\omega_{s}) > -1$) and $\omega_{mp} >
\omega_{p}$ (or $\mu_{2}(\omega_{s}) < -1$).  In the former case, there will be
at least two transmission resonances at a given $\omega$ for a deviation of $\epsilon_2$ from the
perfect lens condition.  In fact, for small enough negative deviation
$\epsilon_{2} = -1 -\delta$, ($\delta > 0$, but small), the first condition can
be satisfied for three $k_x$, and upto four transmission resonances are
possible.  In the latter case of $\mu_{2} < -1$, there will be at least one
transmitted resonance that can be excited at any $\omega$.  For small enough
positive deviation $\epsilon_{2} = -1 +\delta$ , ($\delta > 0$, but small), the
second condition can be satisfied for two $k_{x}$ and upto three transmission
resonances are possible.  This becomes clear from figure~5(b) and
(d), and the transmission resonances are shown in figure~6(a)
and (c).  Mathematically, the reason for this behaviour can be explained as
follows.  For $k_{x}\rightarrow \infty$, $\omega \rightarrow \omega_{s}\pm
\delta$ (where $\delta>0,$ and $\rightarrow 0$), the $\pm$ sign being chosen
depending on whether $\mu_{2} < -1$ or $\mu_{2} > -1$.  This causes either the
lower or the upper branch respectively to intersect the $\omega = \omega_{s}$
line at some point.  Physically, this is because the behaviour of both the
symmetric and the antisymmetric modes, at large $k_x$, have to tend to the
uncoupled plasmon dispersion for a single surface.  For positive deviation in
$\epsilon_2$ and large enough $k_{0}d$, it is possible that no surface mode
can be excited in either case, but the required amplification of the evanescent
waves also does not result.

[Insert figure 6 about here ] \\

In figure~6, we show the transfer function from the object plane
to the image plane across  a slab of negative
dielectric constant for the  deviations of the various parameters from the
perfect lens condition.  In figure~6(a), we show the transmission
for $\mu_{2} < 1$.  Most values of $\epsilon_{2}$ give a single resonance.  But
for the case of small positive deviation ($\epsilon_{2}= -0.99998$), three
resonances arise, the latter two occur very close together and not resolved in
the graph.  We note in this case that even a small amount of absorption removes
all traces of the resonances at large $k_x$.  In the case of the P-polarized
light and for $\mu_{2} < 1$, the effect of the deviation in $\mu$ is much lesser
than that for $\epsilon$ (real or imaginary).  figure~6(b) shows
the transmission for $\mu_{2} > -1$.  There are now at least two peaks, the one
which was absent in the other case being very close to $k_{x} = k_{0}$.  In
figure~6(c), we show the transmission for $\mu_{2} = 1$.  The
additional transmission resonances at large $k_{x}$ for small negative deviation
in $\epsilon_{2}$ ($=-1.001$) are clearly visible (expanded and shown in
figure~6(d)).  Again, we note that non-zero absorption in the slab
prevents the divergences at the resonances.  But the resonance at $k_{x}$ close
to $k_0$ is not as strongly damped by absorption.

The transmission resonances will cause the Fourier components at that spatial
frequency to be disproportionately represented in the image and will introduce
`noise' in the image.  The slab will also act as a `low-pass' filter for the
spatial frequencies because of  the deviation (in the real or the imaginary 
parts of $\epsilon_2$ and $\mu_2$) from the perfect lens condition and
limits the image resolution \cite{smithprep}.  However, any finite amount of
absorption in the slab will not allow the transmission to diverge.  When
the effects of absorption is included in the calculation, the
transmitted intensity at the resonances will be large but finite, and we expect
that the effects of the resonances will be softened.  But absorption also
increases the deviation from the perfect lens conditions and will limit the 
range of spatial frequencies with effective amplification. Hence absorption
by itself  would not
cause a fundamental improvement in the image resolution.

The problem in using the slab of silver as a perfect-lens, even for P-polarized
light, is that the imaging is severely affected by the effects of 
retardation arising from large deviation in the
magnetic permeability ($\mu = 1$ everywhere) from the perfect lens condition
($\mu_{2} = -\mu_{3}$).  As it happens, this large deviation of $\mu$ severely
restricts the range of Fourier components which are effectively amplified in the
silver film.  In fact, the effect of absorption, other than to soften the
effects of the transmission resonances, turns out to be small in comparision.
For the case of a lossless (with a very small absorption) symmetric slab with
negative dielecric constant ($\epsilon_{2} = -1$, $\mu = 1$), the plasmon
frequency is $\hbar \omega_{sp} = 3.48 eV$, and $k_{0} = 1.77 X 10^{-2}$
nm$^{-1}$.  As can be seen in figure~7(a), the effects of
retardation become severe and the image of a pair of slits is completely swamped
by the `noise' and a large peak develops in-between the images of the slit.  The
positions of the peaks are also different.  However, the presence of absorption
softens all the singular behaviour at the resonances and we end up with a good
image, in spite of the effects of retardation which is now mainly to reduce the
intensity of the image.  This is shown in figure~7(b) for the case
of the symmetric lossy slab of silver in air.  The central lobe is very small
now and the positions of the peaks are slightly shifted.  Although there is
significant distortion of the image,  the main features
of the image are seen to be unaltered.  Thus, absorption has an interesting role
of limiting the resolution, and yet being vital for the image formation.

[Insert figure 7 about here ] \\

Now we will consider the effects of retardation on the asymmetric, lossy,
negative dielectric lens, with a positive magnetic permeability($\mu_{2}=+1$).  
In this case,
the dispersion for the surface mode looks like the lower branch in 
figure~5, but with a different value of 
$\omega_{s}=\omega_{p}/\sqrt{1+\epsilon_{3}}$.   The large  $\epsilon_{3}$, 
however, amplifies the effects of the large deviation of the magnetic 
permeability ($\mu
=1 $ everywhere) from the perfect lens conditions.  This drastically reduces
the range of Fourier frequencies for which effective amplification is possible.
This effect is, however, offset by the beneficial and larger effect
caused by a reduced optical frequency ($\hbar \omega = 2 eV$) for the surface
plasmon excitation at the silver-silicon interface.  Thus, the effects of
retardation are much reduced and we are closer to the electrostatic limit in
this case.  The image obtained for the asymmetric lossy lens with silicon on
the image side ($k_0 = 1.02X 10^{-2}$ nm$^{-1}$ corresponding to $\hbar \omega
= 2 eV$) is shown in figure~7(c).  The effects of retardation on 
the image in this case
are seen to be much smaller and the image is not `noisy', although the
intensity is much reduced due to the lower transmission, when compared to the
electrostatic case.  Note that we are able to resolve sub-wavelength features
in this case at about $\lambda /10$ lengthscale. This should be compared to the
case of the symmetric lens where a resolution of only about $\lambda/4$ is
obtained, but at a smaller wavelength.

[Insert figure 8 about here ] \\

	 Finally, we explore the possibility that a deviation of $\epsilon_2$
from the perfect lens condition may be used to compensate for the deviation of
$\mu_2$ in order to obtain a better image.  We show the transmission across the
asymmetric slab in figure~8 for different $\epsilon_2$,
while keeping $\epsilon_{1}=1$ for air and $\epsilon_{3}=+15$ for silicon fixed.
This can again be achieved by tuning the frequency of light and using the
material dispersion of silver.  We note that the slab mode can be excited
by making $\epsilon_{2}$ more negative ($=-\epsilon_{3} -\delta$, where
$\delta>0$), but the transmission remains finite due to the finite absorption
even at the resonant $k_x$.  The transmission is increased within a finite
window of $k_x$ for such a deviation, even though the transmission decays much
faster at higher values of $k_{x} \gg k_{0}$.  This is clearly seen in
figure~8 for the cases $\epsilon_{2}(\hbar\omega=1.89 eV)
=-17+i0.4$ and $\epsilon_{2}(\hbar\omega=1.85 eV) =-18+i0.4$.  A deviation of
$\epsilon_2$ on the other side (making it more positive) 
just reduces the transmission even further.  
Thus, we can obtain some extra transmission at lower (but 
sub-wavelength) spatial frequencies at the cost of reduced transmission at 
higher spatial frequencies.  The enhancement in the image features obtained by 
such a compensation for the deviation in $\mu$ by changing $\epsilon_{2}$ from 
the perfect lens condition is shown in figure~7(c) by the dashed
curve.  The enhanced resolution is clearly visible.

\section{Conclusions} 

In conclusion, we have shown that a slab of negative refractive index imbedded
between two media of different (positive) refractive index also behaves as a
near-perfect lens whose performance is limited by only the effects of absorption
and retardation. It is only necessary that {\it one} of the interfaces satisfies
the conditions for the existance of a surface mode at the operating frequency.   We have shown that the asymmetry can be used to enhance the
resolution of the lens against the effects of absorption in the slab.  In
particular, we examined the case of a negative dielectric lens consisting of a
slab of silver with air on one side and any other dielectric medium on the
other.  We found that retardation effects due to a finite frequency of light
and the mismatch in the magnetic permeability ($\mu =1 $ everywhere) limit the
image resolution severely, and also results in the excitation of slab
resonances that degrade the performance of the lens.  Absorption alleviates the
effects of the sharp resonances to some extent, and indeed, is vital to the
imaging itself.  The reduced frequency of light corresponding to 
the surface plasmon excitation at the
interface between silver and high-dielectric-constant medium can reduce the
adverse effects of the retardation drastically. We have shown that
sub-wavelength imaging at optical frequencies ($\hbar \omega \simeq 2.0 eV$) is
possible by our asymmetric lens consisting of a slab of silver bounded by air
on one side and silicon or GaAs on the other, and we obtain a spatial
resolution of about 60-70 nm with this system. 

	The possibility of having a medium to support a thin film of silver
permits the construction of a nanoscopic mechanically rugged lens for optical
near-field imaging.  It also admits the possibility of an integrated detection
system by placing, for example, an array of quantum dots at the image plane
inside the GaAs (medium-3).  It would also be easier to achieve better surface
uniformities , which will eventually limit the performance of the lens.  One
more advantage is that the operating frequency is reduced from $\hbar\omega =
3.48 eV$ in the UV for the symmetric silver slab in air to $\hbar\omega \simeq
2 eV$ in the visible for the asymmetric lens.  The advantages of this reduction
of operating frequency from the UV to the visible for optical instrumentation
are obvious.  The one comparative disadvantage of the asymmetric lens is a
non-zero reflection coefficient which can introduce artifacts in the image.This
is because the reflected field can now be multiply scattered between the source
and the lens to disturb the object field.  This is, however, a problem with the
current methods of SNOM as well.  It might become necessary to account for the
reflection in order to obtain good images.

\section*{Acknowledgments}
The authors would like to acknowledge the support from DoD/ONR MURI grant 
N00014-01-1-0803.

\newpage
\begin{center} {\bf Figure captions} \end{center}
\noindent Figure 1: The asymmetric lens system consisting of a slab of
negative refractive index
bounded by air on one side and any positive medium on the
other.  The object in medium-1 at a distance $d/2$ from the silver surface is
focussed inside the medium-3.  The partial reflection/transmission coefficients
across the boundaries are shown. \\

\noindent Figure 2: Schematic of the field strength distribution for i
(a)~The asymmetric slab with
the surface plasmon excited at the air-silver interface, (b)~The asymmetric
slab with the
surface plasmon excited at the interface between silver and the higher
dielectric constant medium, (c)~The symmetric slab with air on both sides. The
field strength at the image plane is the same as the object plane in this case.\\

\noindent Figure 3: The images (intensity at the image plane in arbitrary units)
 obtained in the electro-static approximation for the
favourable and unfavourable configurations showing the effects of absorption
in the slab on the image resolution.  The positions of the slits which are
imaged are shown by dotted lines. The dielectric constants of the media are
shown within the brackets in the legends.
(a) Enhancement of the image resolution
caused by the asymmetric lens when $\epsilon_{2} = -\epsilon_{3} +i 0.4$, {\it
i.e.}, the favourable configuration.
The slits (20nm wide and centre-centre
spacing of 80 nm)  are just resolved for the symmetric case (air), but are
well resolved for large $\epsilon_{3}=14.9$ of Silicon.
(b) Degradation of the image resolution with increasing
dielectric constant of medium-3, when $\epsilon_{2} = -1 +i 0.4$, i.e., the
unfavourable configuration.  The slits, 30nm wide (larger compared to (a)) with
a centre-centre spacing of 120nm (more compared to (a)),
are barely resolved for the case of Silicon. \\

\noindent Figure 4: The dispersion for the surface plasmon at an interface 
between a negative and a positive medium when the effects of
dispersion are included. The frequency axis is scaled with respect to the bulk
plasmon frequency ($\omega_{p}$) and the $k_x$ axis is scaled with respect to
$k_{p} = w_{p}/c$. The light line $\omega = ck_x$ and the
$\omega=\omega_{s}=\omega_{p}/\sqrt{2}$ line are given by the dotted lines. The
dispersion inside the infinite negative medium, $\omega =
ck_{x}/\sqrt{\epsilon_{2}\mu_{2}}$ with $\omega_{mp}=2\omega_{p}$ is shown by
the ---$\cdot$--- curve (1). \\

\noindent Figure 5:  
The dispersion for the coupled surface plasmons when the effects of
retardation are included. The frequency axis is scaled with respect to the bulk
plasmon frequency ($\omega_{p}$) and the $k_x$ axis is scaled with respect to
$k_{p} = w_{p}/c$. The light line $\omega = ck_x$ and the
$\omega=\omega_{p}/\sqrt{2}$ line are given by the dotted lines.
(a)~$\mu_{2}=+1.0$, (b)~behaviour in (a) at larger $k_x$.
(c)~$\mu = 1- \omega_{mp}^{2}/ \omega^{2}$ with  $\omega_{mp}=2\omega_{p}$. The
dispersion in the infinite negative medium, $\omega =
ck_{x}/\sqrt{\epsilon_{2}\mu_{2}}$, is shown by the ---$\cdot$--- curve. The part
of the lower branch with the $\diamond$ symbol shows the non-radiative light
modes within the slab.
(d)~behaviour at larger $k_x$ in (c). \\

\noindent Figure 6: 
The absolute value of the) transmitted field at the image plane
$\vert T \exp(ik_{z}^{(1)}d)\vert$, for
the symmetric slab  as a function of $k_x$, showing the resonances caused by
deviations in $\epsilon$ and $\mu$ from the perfect lens conditions. The
$\epsilon$ and $\mu$ indicated in the graphs are for the material of the slab.
It is assumed $\epsilon = 1$ and $\mu=1$ outside the slab. The X-axis is scaled
with respect to the free-space propagation vector $k_0$.\\

\noindent Figure 7:
Effects of retardation on the images. $\mu = +1.0$ everywhere. The
thickness of the slab $d=40 nm$. The object is a pair of slits of width 20 nm,
placed apart by 100 nm shown by the dotted lines. (a) the symmetric lossless
lens. (b) The symmetric lossy silver lens with ($\hbar\omega = 3.48 eV$) (c)
The asymmetric lossy silver slab with Silicon as medium-3
[$\epsilon_{3}(\hbar\omega = 2 eV) \simeq 15$]. The dashed line is for the case
when a deviation in $\epsilon_{2} =-18.0+i0.4$ compensates for the deviation in
$\mu$ from the perfect lens conditions.\\

\noindent Figure 8: 
The absolute value of the transmitted field at the image plane
across the asymmetric slab for different $\epsilon_2$.
$\epsilon_{1}=1.0$ and $\epsilon_{3}=15.0$ are kept fixed. 

\newpage

\begin{figure}
\vspace{4cm}
\epsfxsize=450pt
\begin{center}{\mbox{\epsffile{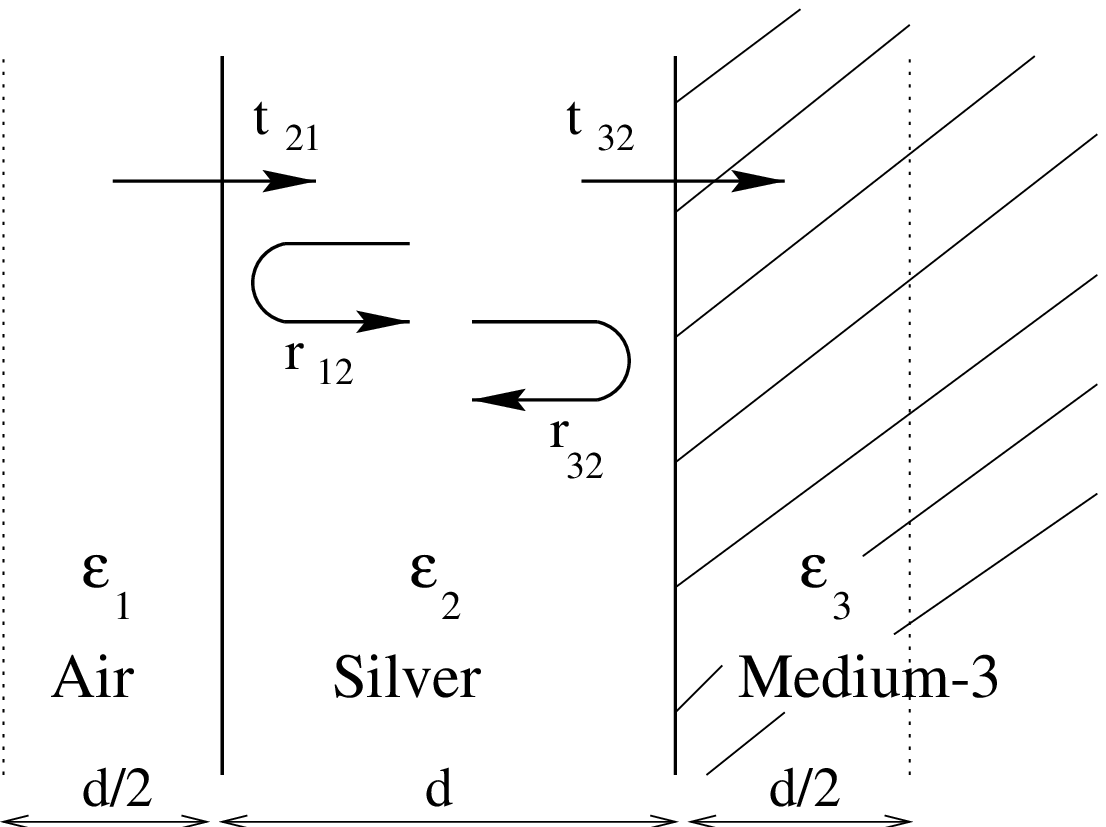}}} \end{center}
\end{figure}

\newpage

\begin{figure}
\vspace{4cm}
\epsfxsize=450pt
\begin{center}{\mbox{\epsffile{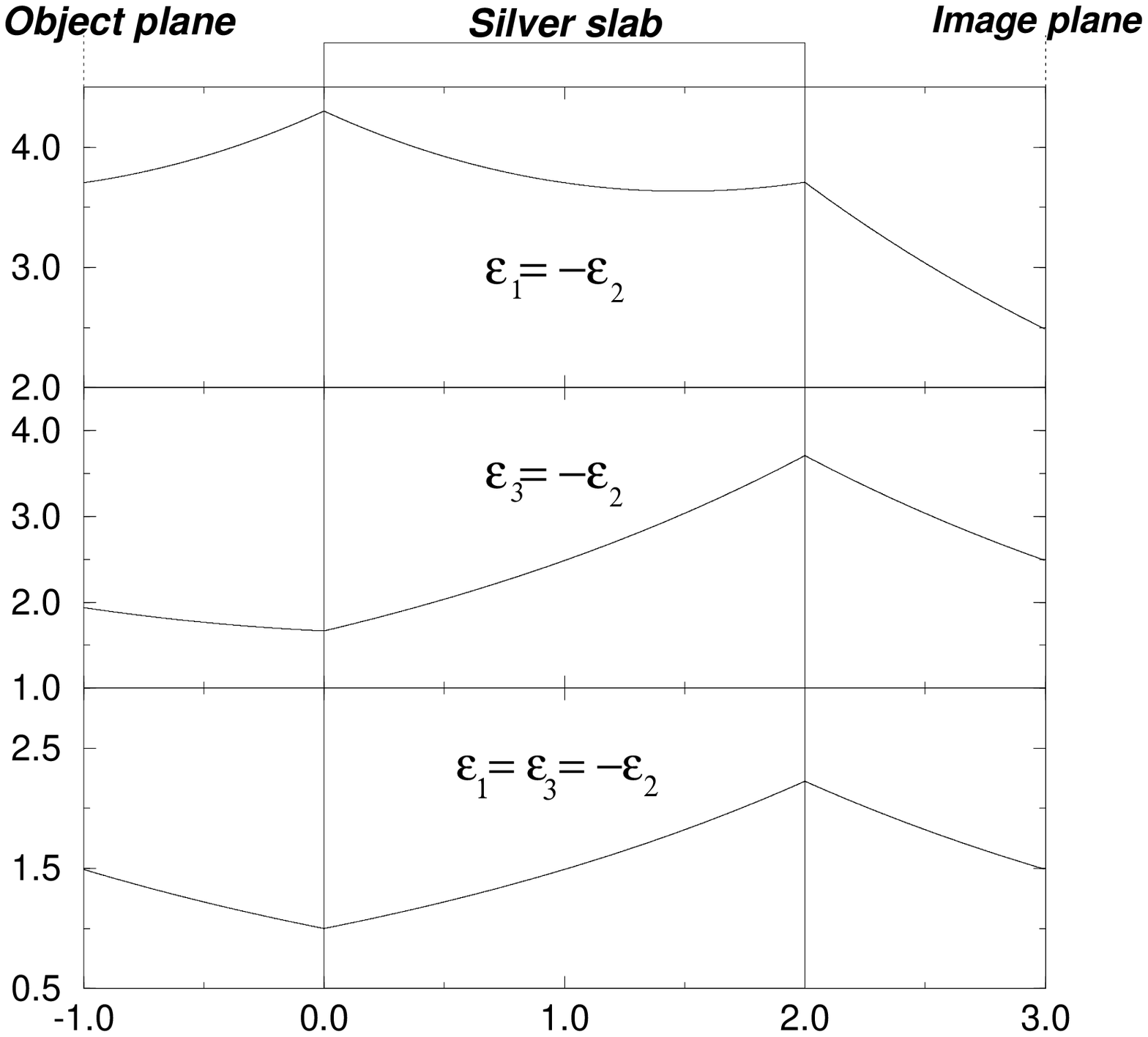}}} \end{center}
\end{figure}

\begin{figure}
\vspace{4cm}
\epsfxsize=450pt
\begin{center}{\mbox{\epsffile{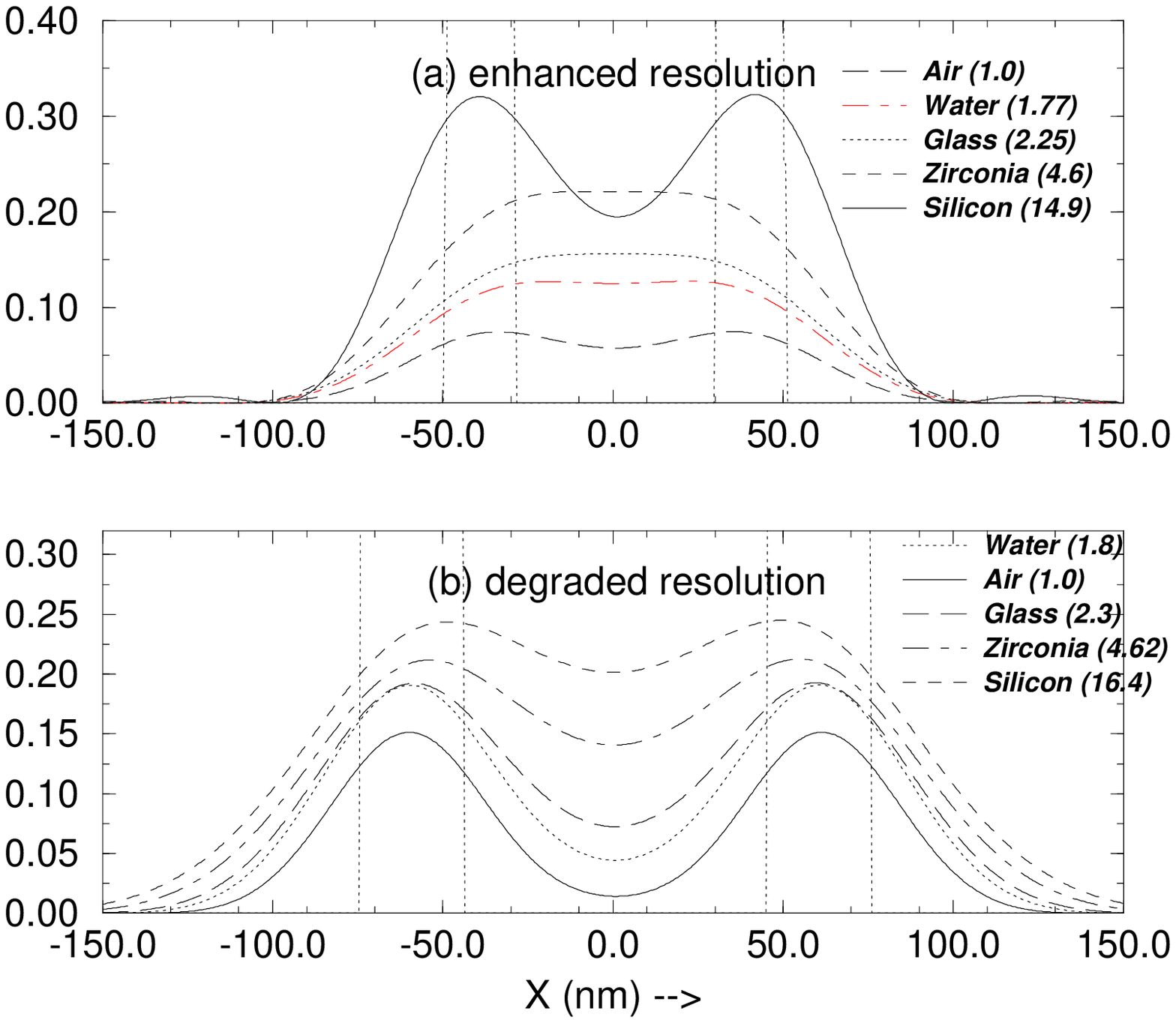}}} \end{center}
\end{figure}

\newpage
\begin{figure}
\vspace{4cm}
\epsfxsize=450pt
\begin{center}{\mbox{\epsffile{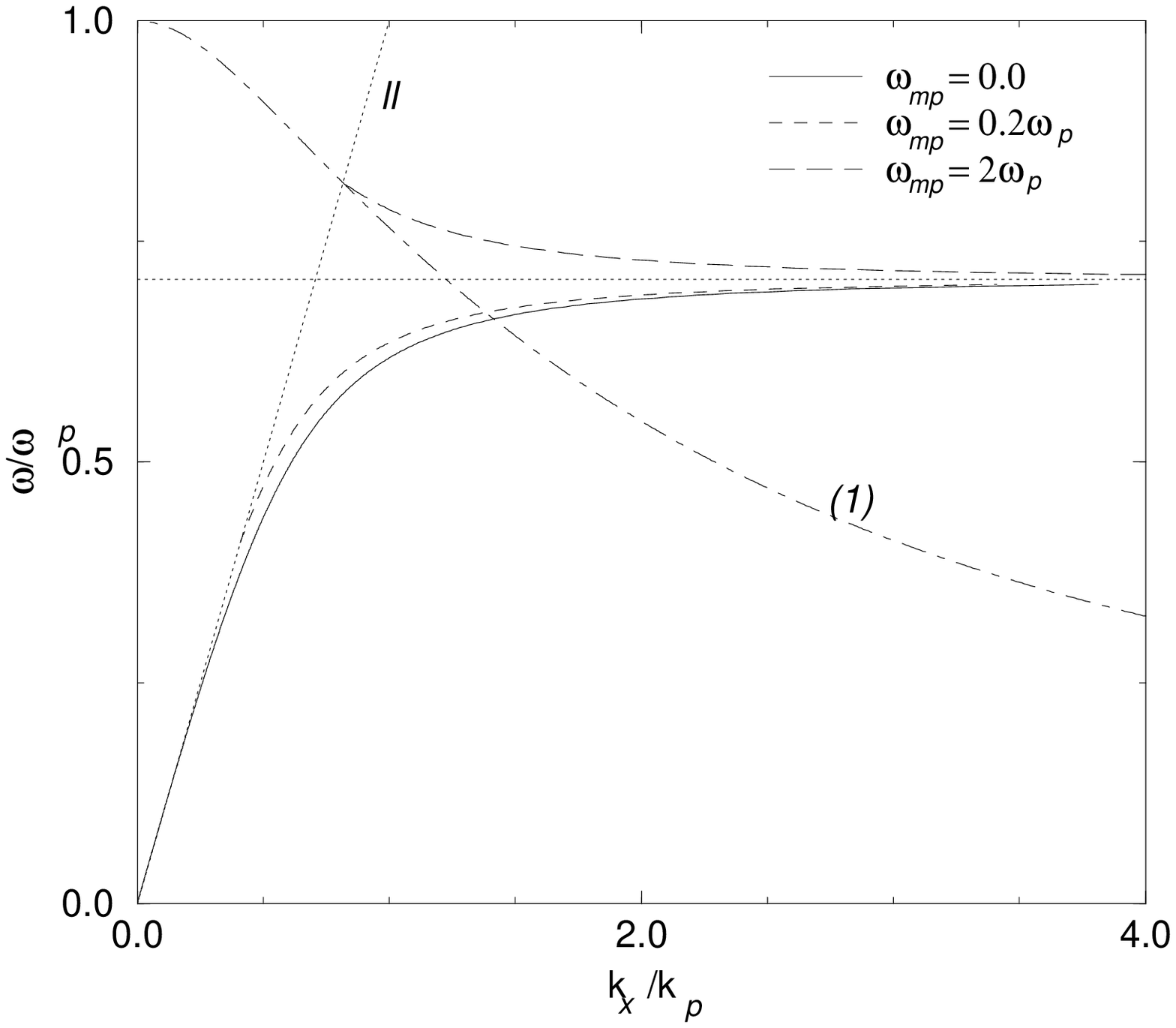}}} \end{center}
\end{figure}

\newpage
\begin{figure}
\vspace{4cm}
\epsfxsize=450pt
\begin{center}{\mbox{\epsffile{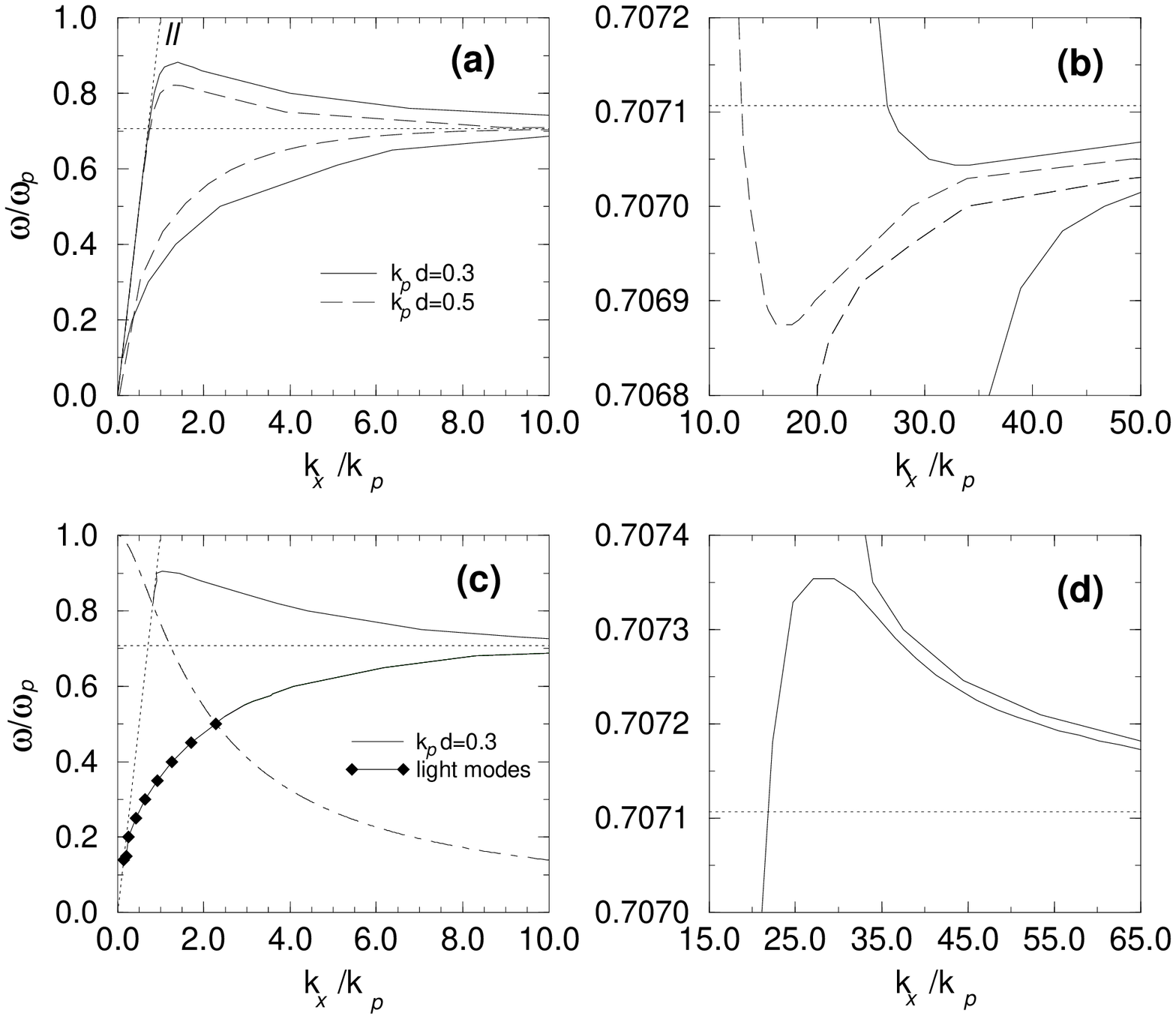}}} \end{center}
\end{figure}

\newpage
\begin{figure}
\vspace{4cm}
\epsfxsize=450pt
\begin{center}{\mbox{\epsffile{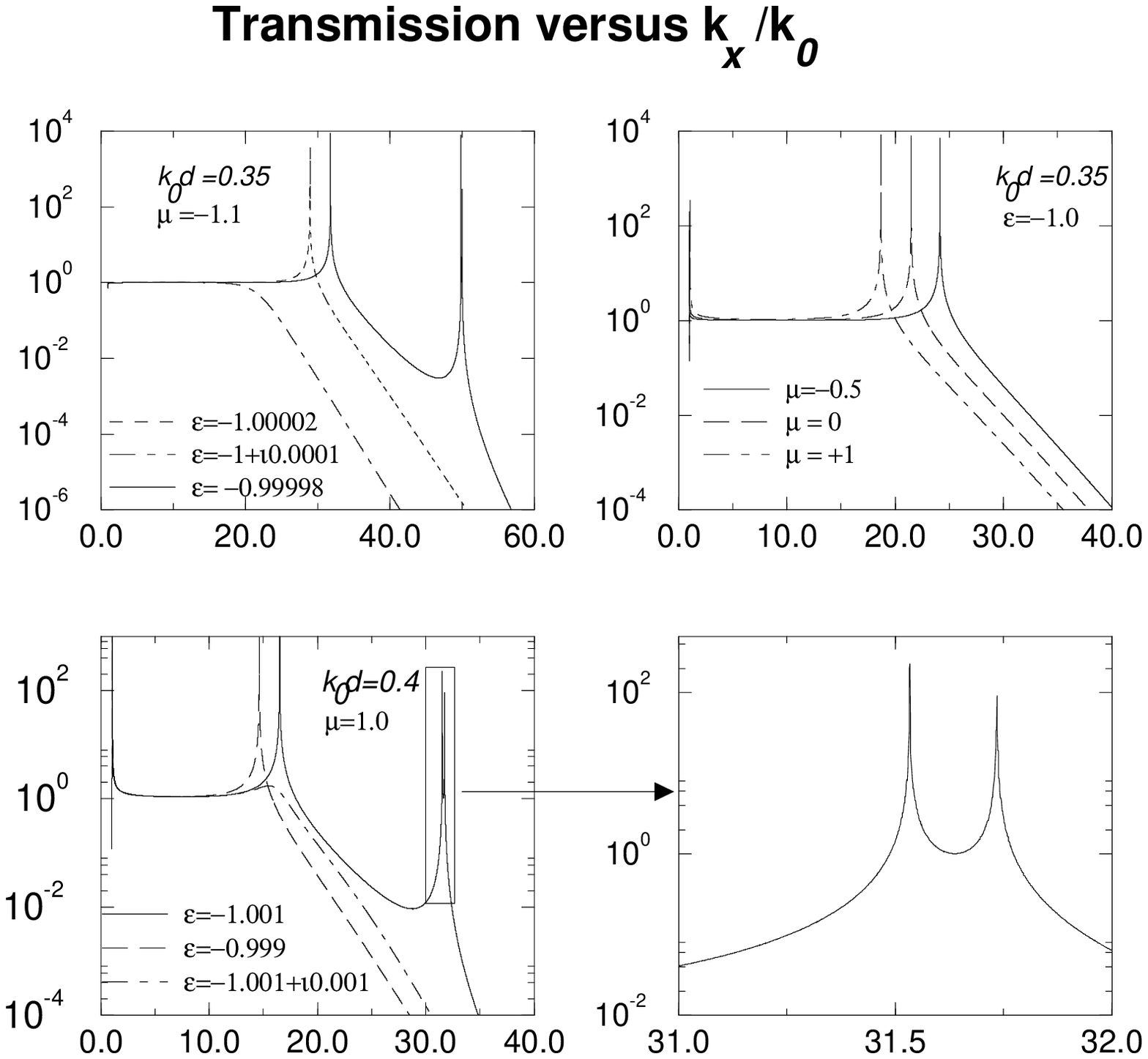}}} \end{center}
\end{figure}

\newpage
\begin{figure}
\vspace{4cm}
\epsfxsize=450pt
\begin{center}{\mbox{\epsffile{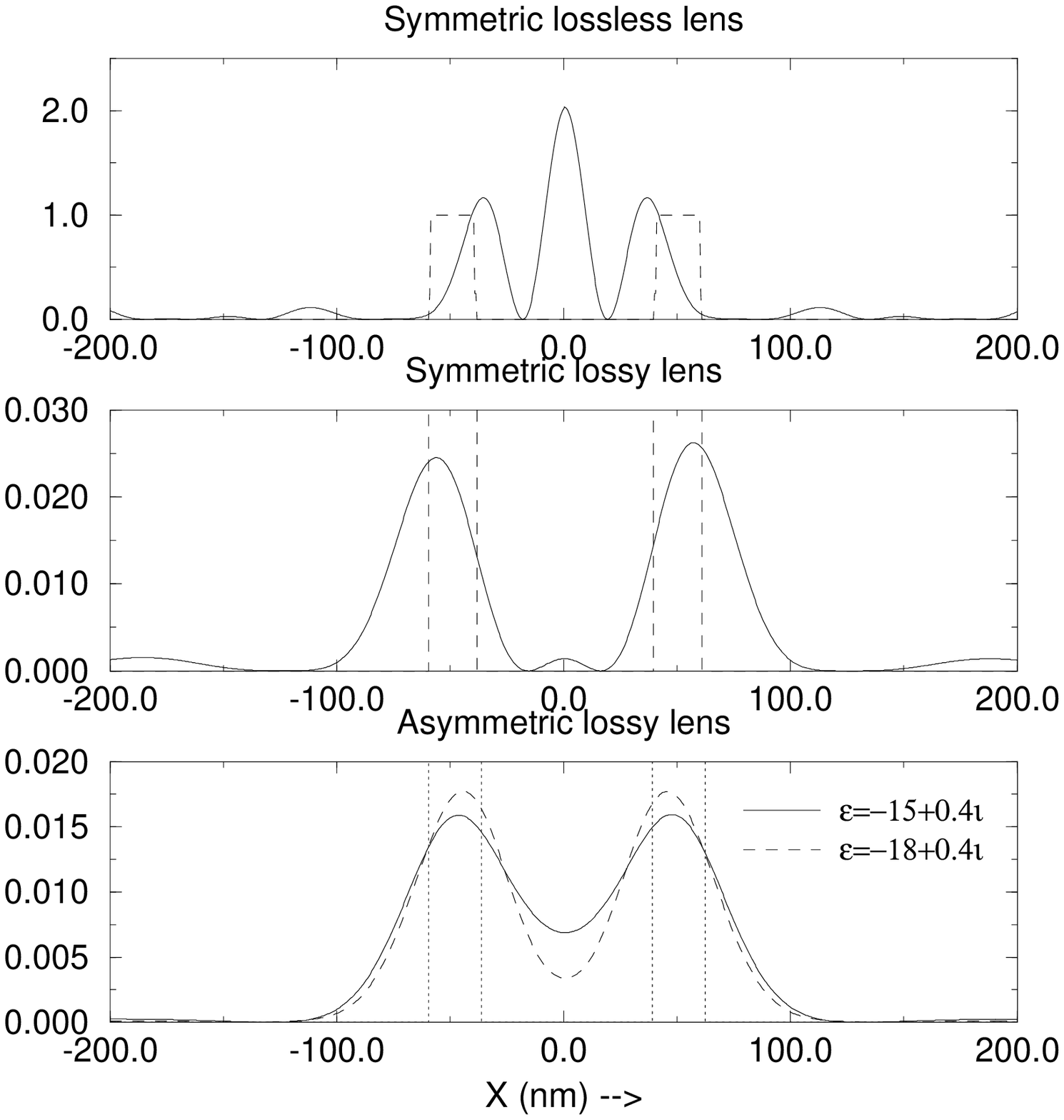}}} \end{center}
\end{figure}

\newpage
\begin{figure}
\vspace{4cm}
\epsfxsize=450pt
\begin{center}{\mbox{\epsffile{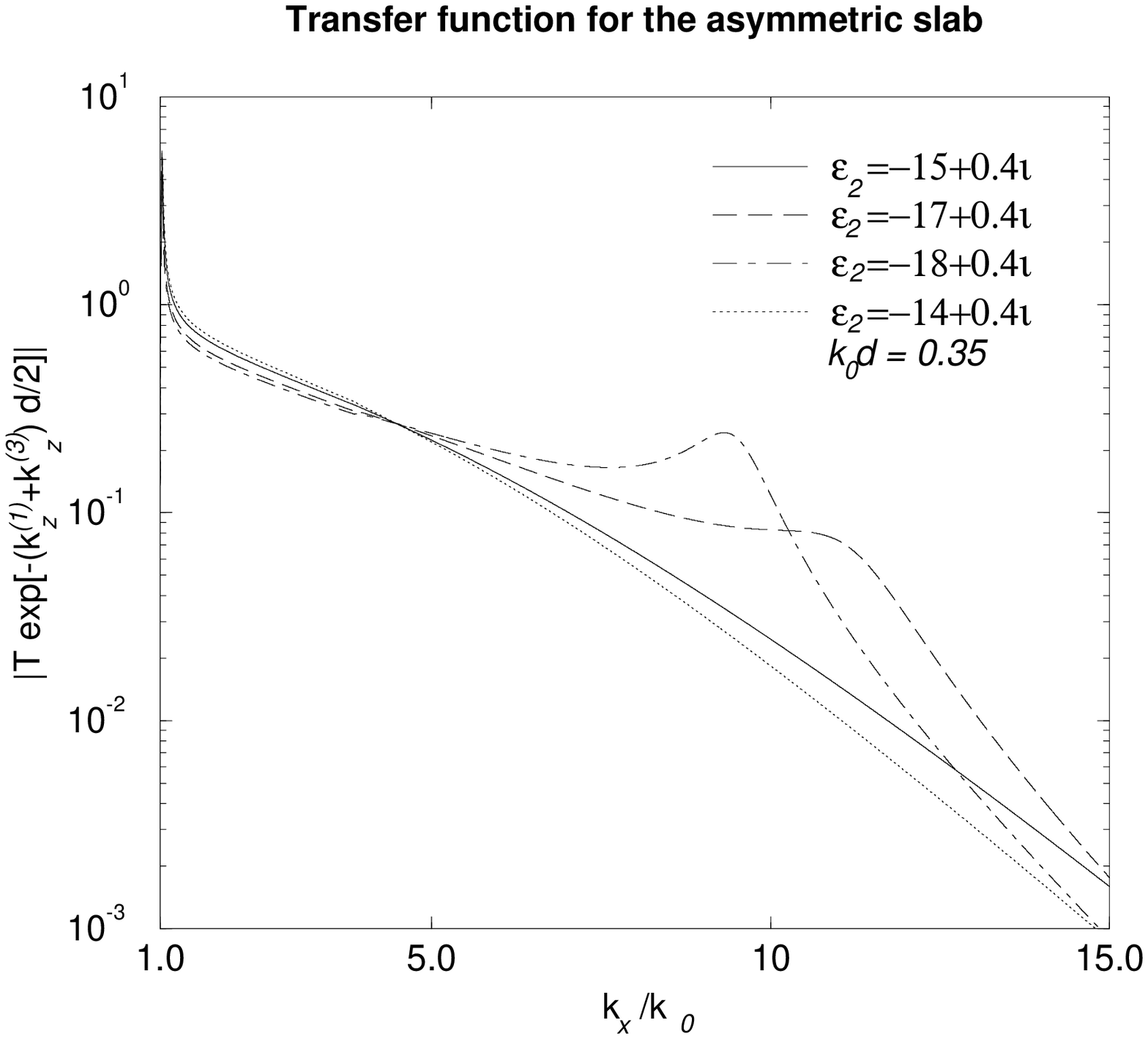}}} \end{center}
\end{figure}
\end{document}